\documentclass[letterpaper, 10 pt, conference]{ieeeconf}
\IEEEoverridecommandlockouts 
\usepackage{amsmath,amssymb,amsfonts}
\usepackage{algorithmic}
\usepackage{graphicx}
\usepackage{textcomp}
\usepackage{xcolor}
\usepackage[style=numeric-comp,sorting=none]{biblatex}
\usepackage{hyperref}
\usepackage{tikz}
\def\BibTeX{{\rm B\kern-.05em{\sc i\kern-.025em b}\kern-.08em
    T\kern-.1667em\lower.7ex\hbox{E}\kern-.125emX}}

\hypersetup{
    bookmarks=true,         
    unicode=false,          
    pdftoolbar=true,        
    pdfmenubar=true,        
    pdffitwindow=false,     
    pdftitle={Unified  Mobility  Estimation  Model},    
    pdfauthor={David Ziegler, Johannes Betz, Markus Lienkamp},     
    pdfsubject={},   
    pdfcreator={David Ziegler},   
    pdfproducer={},  
    pdfkeywords={Mobility,} {Estimation,} {Probabilistic}, 
    pdfnewwindow=true,      
    colorlinks=false,       
    linkcolor=red,          
    citecolor=green,        
    filecolor=magenta,      
    urlcolor=cyan           
}

\newcommand\copyrighttext{%
  \footnotesize \textcopyright 2021 IEEE. Personal use of this material is permitted.
  Permission from IEEE must be obtained for all other uses, in any current or future
  media, including reprinting/republishing this material for advertising or promotional
  purposes, creating new collective works, for resale or redistribution to servers or
  lists, or reuse of any copyrighted component of this work in other works.
  DOI: \href{https://ieeexplore.ieee.org/document/9564453}{10.1109/ITSC48978.2021.9564453}}

\newcommand\copyrightnotice{%
\begin{tikzpicture}[remember picture,overlay]
\node[anchor=south,yshift=10pt] at (current page.south) {\fbox{\parbox{\dimexpr\textwidth-\fboxsep-\fboxrule\relax}{\copyrighttext}}};
\end{tikzpicture}%
}

\title{\LARGE \bf
Unified Mobility Estimation Model
}

\author{David Ziegler$^{1}$, Johannes Betz$^{2}$, Markus Lienkamp$^{3}$
\thanks{$^{1}$David Ziegler is with the Institute of Automotive Technology, Technical University Munich, Germany, ORCID: 0000-0003-3494-3034}%
\thanks{$^{2}$Johannes Betz is with the Dept. of Electrical and Systems Engineering, University of Pennsylvania, USA, ORCID: 0000-0001-9197-2849}%
\thanks{$^{1}$Markus Lienkamp is with the Institute of Automotive Technology, Technical University Munich, Germany, ORCID: 0000-0002-9263-5323}%
}

\begin{document}
\maketitle
\copyrightnotice
\thispagestyle{empty}
\pagestyle{empty}

\begin{abstract}
In literature, scientists describe human mobility in a range of granularities by several different models. Using frameworks like MATSIM, VehiLux, or Sumo, they often derive individual human movement indicators in their most detail. However, such agent-based models tend to be difficult and require much information and computational power to correctly predict the commutation behavior of large mobility systems. Mobility information can be costly and researchers often cannot acquire it dynamically over large areas, which leads to a lack of adequate calibration parameters, rendering the easy and spontaneous prediction of mobility in additional areas impossible. This paper targets this problem and represents a concept that combines multiple substantial mobility theorems formulated in recent years to reduce the amount of required information compared to existing simulations. Our concept also targets computational expenses and aims to reduce them to enable a global prediction of mobility. Inspired by methods from other domains, the core idea of the conceptional innovation can be compared to weather models, which predict weather on a large scale, on an adequate level of regional information (airspeed, air pressure, etc.), but without the detailed movement information of each air atom and its particular simulation.

\end{abstract}

\section{Introduction} \label{section:Introduction}
\begin{figure}[h]
\centerline{\includegraphics[width=9cm]{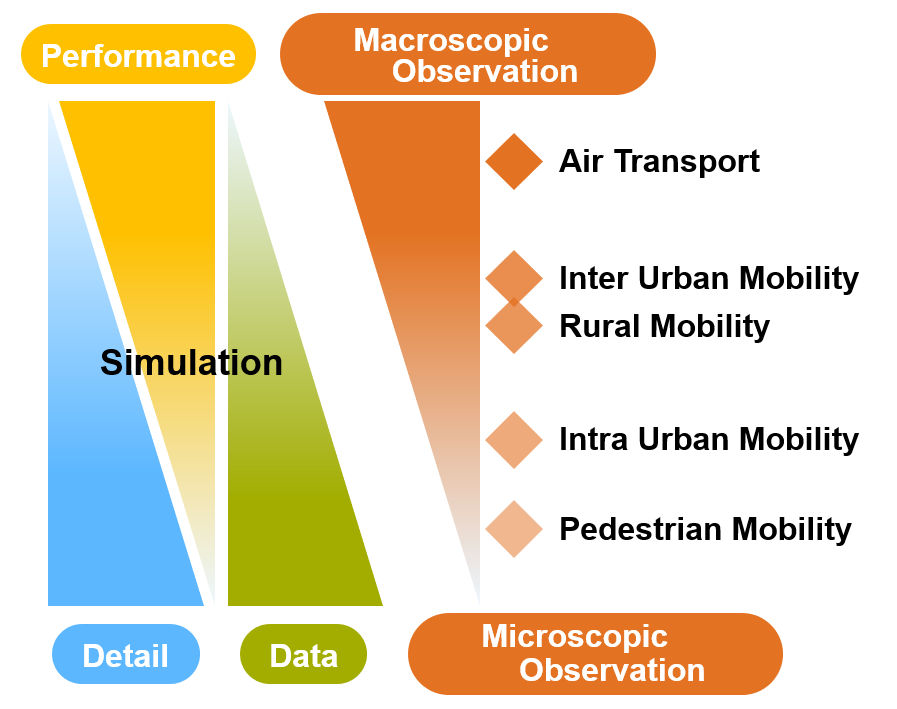}}
\caption{Tendencies of simulation detail, computational performance and data requirements of mobility simulations depending on their scope of observation.}
\label{fig:1}
\end{figure}

Mobility in this paper means spatial mobility, especially the movement of persons and goods in geographical space. It can be described with different levels of detail and granularities. In literature, natural movements are generally classified according to the distances they overcome within one trip, referred to Barbosa et al. \cite{Barbosa.2018} as jump lengths. Categorizing mobility with this metric, air transport \& airport management \cite{Guimer.2004, Cook.2016} involves the biggest jump lengths, whereas inter-urban mobility \cite{Ren.2014, Ravenstein.1885} and rural mobility \cite{Ravenstein.1885, FilippoSimini.2012, Pappalardo.2015} are observed on a smaller regional scale, which involves smaller jump lengths. Thus, together with intra-urban mobility \cite{Hanson.1980, Zheng.2011}, they are often represented in microscopic multi-agent simulations \cite{Horni.2016, Kumar.2019, Hager.2015} that aim to describe mobility behavior precisely for a region's road-user subset \cite{BernhardLuger.2017}. Mobility data of the collective is subsequently extrapolated from the results of the subset simulations. Inspired by particle mass flows, pedestrian simulations describe mobility behavior on a pedestrian level \cite{Karamouzas.2014} to predict forces acting on individuals or the spread of pathogens during a pandemic \cite{Eubank.2004}. Commutation flows can be described either statically in time or dynamically \cite{Zanin.2013}. However, literature frequently shows that the smaller the scope of observation, the more detailed models dominate. This stems from the recent tendency in research to derive knowledge from dynamic multi-agent interactions that require a precise simulation of individual mobility behavior. With increasing detail, there comes increasing complexity, which requires more information to adequately calibrate the models. Computational expenses usually increase with the degree of detail, which renders an extensive simulation on a large scale inappropriate \cite{Balmer.2008, Balmer.2009}. \autoref{fig:1} illustrates from the macroscopic to the microscopic magnitude how the simulation performance tends to depend on the degree of detail and the amount of data the underlying model requires.   

Focusing on inter-urban, intra-urban and rural mobility, the state of the art displays significant improvements in algorithms for reducing the computational expenses of multi-agent simulations \cite{Zhuge.2019,Waraich.2015,Charypar.2006, Lefebvre.2007}. The synthesis of multi-agent sets having a realistic socio-ethnic composition for a certain target area has been well researched in \cite{BernhardLuger.2017, Moeckel.2003, Moreno.2018}. Although these improvements significantly increase computational speed and decrease synthesis complexity, they still have three major downsides that owe their existence to the nature of multi-agent simulation:
\begin{enumerate}
	\item[1.] To describe the agents’ mobility behavior realistically, demographic, cultural, and behavioral information of the society in the target area is required \cite{BernhardLuger.2017, Balmer.2009}.
    \item[2.] By increasing the simulated area, the number of agents and computational nodes also increases, which usually leads to a non-linear increase in computational complexity \cite{Balmer.2009}.
    \item[3.] Due to the aforementioned disadvantages, only a certain proportion of local inhabitants can in most cases be effectively simulated for a region. This causes discrepancies between the simulation results and real-world observations. Furthermore, information that is needed to calibrate the models is often not available in the required detail which hinders a spontaneous and flexible simulation of additional target areas.
\end{enumerate}

Considering the disadvantages of micro-scale simulations, it is necessary to explore alternative mobility concepts. These concepts should be less detailed and provide sufficient  global mobility indicators while avoiding significant losses in accuracy. Therefore, this paper proposes a new concept for modelling mobility in inter-urban and rural areas intending to make mobility simulations quicker, easier, and more flexible by limiting the amount of information and computational expense agent-based frameworks typically require. Our model combines existing concepts from literature to simplify mobility simulations, especially for large target areas.

\section{Related Work} \label{section:Related_Work}

This paper incorporates existing mobility methods from literature focusing on the fundamental laws of mobility established by commutation proportions and constraints. As the paper concentrates on rural and inter-urban mobility, literature focusing on macroscopic observations such as passenger flows between airports or microscopic observations like individual pedestrian mobility is not considered. The paper also examines daily and weekly routines that constitute the most traffic load in rural and inter-urban areas. Occasional traffic events and transit traffic are thus not considered. The following subsections describe mobility-relevant proportionalities, related behavior models, and scaling effects threaded by the equilibrium approach.

\subsection{Ranked Distributions} \label{section:Ranked_Distributions}
\subsubsection{Proportionality - Zipf’s Law} \label{section:Zipf_Law}
Many assumptions discussed in the following chapters are primarily based on proportions that form distributions. One distribution, named after its discoverer, George Kingsley Zipf, is of particular significance. The Zipf’s Law/distribution \cite{Zipf.1950}, later improved by the rank-size rule \cite{Hinloopen.2006}, relates a value $f_z$ to the corresponding rank z over all observed values $F_Z$.

\begin{equation} 
f_z\left(z\right)\sim\frac{1}{z^\alpha}\ \ |\ f_z\subset F_Z
\end{equation}.

While Zipf’s Law for instance describes the population size of towns according to their population rank \cite{Hinloopen.2006}, it is an integral element of modern publications in multiple domains. According to Zipf, commodity flows follow the Zipf's Law because they underlie the “force of diversification” and the “force of unification” and balance out at some point to a proportion. Further, it is possible to derive an estimation of the goods-flow $w_{ij}$ between human settlements according to their population sizes $(P_i,P_j)$ and distance $r_{ij}$ \cite{Barbosa.2018}:

\begin{equation} 
w_{ij}\sim\frac{P_iP_j}{r_{ij}}
\end{equation}

Similar rules apply to human needs for commutation as most people tend to visit locations in their surroundings repeatedly. According to location visiting ranks $z_v$ which show the locations’ importance in daily life, the visiting frequencies $f_v$ of a location (most simply with $\alpha=1$) can be approximated \cite{Gonzalez.2008}:

\begin{equation} 
f_v(z_v)\sim z_v^{-\alpha}
\end{equation}.

The probability function is derived accordingly \cite{Gonzalez.2008}:

\begin{equation} 
P\left(f_v\right)_{|z_v}\sim z_v^{-\left(1+\frac{1}{\alpha}\right)}
\end{equation}

In general, social and cultural activities are subject to the same fundamental forces of diversification and unification and are therefore also governed by Zipf’s Law. They reflect in the gravity model \cite{Zipf.1946,Erlander.1990, Wills.1986}, the intervening opportunities model \cite{Stouffer.1940} and the radiation \cite{FilippoSimini.2012} model, which will be discussed in the following section.

\subsection{Periodic Commutation Behavior} \label{section:Commutation_Behavior}
For this publication relevant commutation behaviour research was recently published in publications of Zipf \cite{Zipf.1946}, Stouffer \cite{Stouffer.1940}, Simini et al. \cite{FilippoSimini.2012}, Pappalardo et al. \cite{Pappalardo.2015}, Schneider et al. \cite{Schneider.2013}, Domenico et al. \cite{Domenico.08.10.2012}, Song et al. \cite{Song.2010}, González et al. \cite{Gonzalez.2008}, Kölbl and Helbing \cite{Kolbl.2003}. The publications are listed in the order of detail, beginning with general treatments, and discussed in the following paragraphs.

\subsubsection{Gravity Model} \label{section:Gravity_Model}
Zipf describes mobility potentials between geospatial areas with the naturally inspired gravitation model, whereas Stouffer bases his model of mobility flows on the social theorems of intervening opportunities. With a superior view of abstraction that enables predictions in the absence of detailed data, both models aim to predict the commutation between different geospatial areas. Lenormand et al. \cite{Lenormand.2016} compare the performance of the models for short and mid-range trips that are taken daily and conclude that “the distance seems to play a more important role than the number of intervening opportunities“. Therefore, the gravity law with exponential distance decay should be preferred for daily commutation modeling. The rule draws up as follows:
\begin{equation} 
P_{ij}\propto p_i p_j\cdot f_e\left(d_{ij}\right),\ \ i\neq j\ \ |\ \ f_e\left(d_{ij}\right)=e^{-\beta d_{ij}}
\end{equation}

where $P_{ij}$ describes the probability of commuting between two areas with the populations $p_i,p_j$ at a distance $d_{ij}$ to each other.

\subsubsection{Extended Radiation Model} \label{section:Radiation_Model}
In their radiation model, Simini et al. \cite{FilippoSimini.2012} rank destinations according to their importance to the traveler and combine them with the gravity model. Yang et al. \cite{Yang.2014} devises an extension to this model, the 'Extended Radiation Model', which is universally applicable and substitutes the population number with the number of opportunities a destination zone offers (for instance the opportunity to work, to buy foods or to have a haircut). Using the $\alpha$ factor the target area is discretized in $z_{zones}$ quadratic zones with the side length \textit{l}. The model is supplied with the number $n_i$ of opportunities inside the travel origins zone, the number $n_j$ of opportunities inside the travel destination zone, the total number of opportunities $n_{total}$ in the observed region and the number $s$ of opportunities that are passed by en route from zone i to zone j. For travel in inter-urban/rural areas (within a range of: 1 … 65km) the probability $P_{LOC}\left(1\middle|n_i,n_j,s\right)$ of choosing one location inside the destination zone is calculated by:

\begin{gather}
P_{LOC}\left(1\middle|n_i,n_j,s\right)=\frac{P_>\left(n_i+s\right)-P_>(n_i+n_j+s)}{P_>\left(n_i\right)} \nonumber\\
\langle P_>\left(x\right)\rangle=\frac{\frac{1}{1+x^\alpha}-\frac{1}{1+n_{total}^\alpha}}{\frac{1}{1+n_{avg}^\alpha}-\frac{1}{1+n_{total}^\alpha}},\ n_{total}\geq x\geq n_{avg}\label{eq:6}\\
\alpha=\left(\frac{l}{36}\right)^{1.33},\ n_{avg}=\frac{n_{total}}{z_{zones}} \nonumber
\end{gather}

\subsubsection{Returners \& Explorers Model} \label{section:Returners_Explorers}
Pappalardo et al. \cite{Pappalardo.2015} precisely describe mobility based on GPS and GSM data by dividing commuters' into returners' and explorers', according to their commutation behavior. A crucial metric for this classification is the radius of gyration $r_g^{(k)}$ defined over the k most frequently visited locations. The radius calculates the mean square distance between the center of mass $r_{cm}^{\left(k\right)}$ of all visited locations (total number of locations: $N_k$, number of visits an individual location has: $n_i$) and the individual location coordinates $r_i$ as follows:

\begin{equation} 
r_g^{(k)}=\sqrt{\frac{1}{N_k}\sum_{i=1}^{k}n_i\left(r_i-r_{cm}^{\left(k\right)}\right)^2}
\end{equation}

Returners are characterized in the publication by the ratio $s_k$ diverging against 1, while explorers are characterized as diverging against 0:

\begin{equation} 
s_k=\frac{r_g^{\left(k\right)}}{r_g}
\end{equation}

where $r_g$ defines the total radius of gyration over all visited locations.

\subsubsection{Human Mobility Motifs Model} \label{section:Human_Mobility_Model}

\begin{figure}[htbp]
\centerline{\includegraphics[width=9cm]{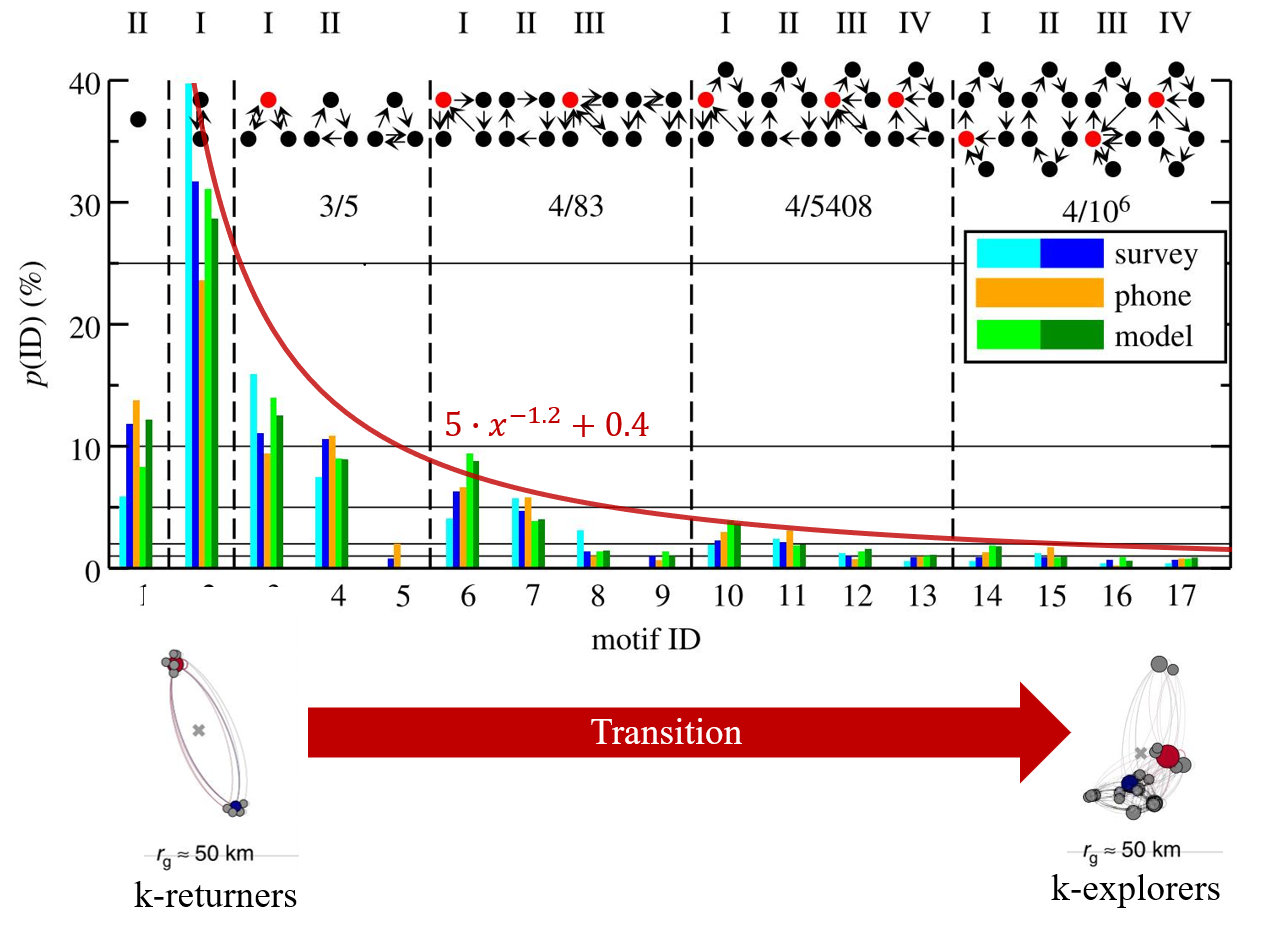}}
\caption{Motifs' transition from returners to explorers, adapted from \cite{Schneider.2013, Pappalardo.2015}.}
\label{fig:2}
\end{figure}

Schneider et al. \cite{Schneider.2013} differentiate commutation flows in distinct detail by classifying the motifs of human mobility using surveys, models, and global GSM data into 17 commutation patterns, covering almost 95\% of the observed travel behavior. These motifs display a transition between the aforementioned ‘returners’ and ‘explorers’ as \autoref{fig:2} illustrates. The explorer probability decreases according to the power law and with an increasing number of considered locations, “k explorers gradually become k returners” \cite{Pappalardo.2015}.

\subsubsection{Socially Confined Interactions \& Preference Locations Model} \label{section:Social_Confined_Model}
De Domenico et al. \cite{Domenico.08.10.2012} explain balancing effects that influence commutation through socially confined interaction. They describe that commuters share travel destinations with friends, colleagues, family and other people they interact daily. This leads to an alignment of individual mobility behavior across different mobility groups. Song et al. \cite{Song.2010} and González et al. \cite{Gonzalez.2008} further mention the daily tendency of travelers to return to frequent destinations, such as their homes, workplaces and other locations.

\subsubsection{Constraint - Physical Travel Energy} \label{section:Physical_Travel_Energy}
It is necessary to also consider single trip and daily distance constrains to describe mobility comprehensively. Accordingly, the Lévy distribution draws up the trip distance distribution $P_{LF}$ of a single trip and its length $r$ \cite{Barbosa.2018}:

\begin{equation} 
P_{LF}\left(r,\mu,c\right)=\sqrt{\frac{c}{2\pi}\ }\frac{e^{-\frac{c}{2\left(r-\mu\right)}}}{\left(x-\mu\right)^\frac{3}{2}\ }
\end{equation}

\begin{figure}[htbp]
\centerline{\includegraphics[width=9cm]{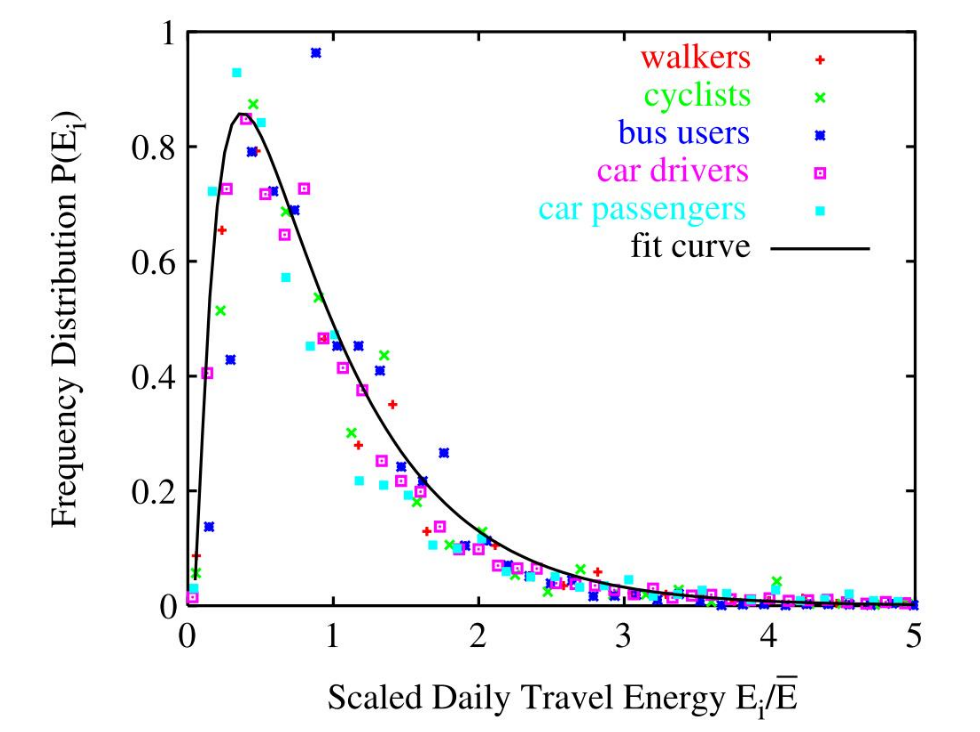}}
\caption{Distribution of personal energy for daily travel, from \cite{Kolbl.2003}.}
\label{fig:3}
\end{figure}

To estimate daily commutation distance constraints, in particular the work of Kölbl and Helbing \cite{Kolbl.2003} needs to be highlighted. Their extensive 25-years-long commutation study showed that the probability to travel a certain distance correlates with the personal physical-human energy a individual spends for the commutation. Accordingly, a person is willing to spend only a certain amount of his/her personal energy on daily travel. This amount of energy virtually underlies a sharply defined canonical distribution (\autoref{fig:3}), whereas on an average day the majority of study participants spent 615 kJ (or below) of their personal energy on commutation. This energy was used, for instance, to sit in a vehicle and drive or to walk a short distance several times a day. It is important to emphasize that the authors only considered personal energy expenditures intentionally. Further, the authors claim that “the daily travel time [$t_i$] is inversely proportional to the rate of energy expended [$ E$]” and has the probability $P_i$ \cite{Kolbl.2003}:

\begin{equation}
P_i\approx\frac{E}{t_i} 
\end{equation}

The aforementioned principles describe commutation behavior of people over a short or medium period of time without the need of multi-agent simulations. To consider the commuters interactions, equilibrated models are required and therefore discussed subsequently.

\subsection{Traffic Equilibrium} \label{section:Traffic_Equilibrium}
The simulation of road participants' routing behavior requires models that accurately resemble real-life routing decisions. Without these models, a realistic estimation of key indicators like fuel consumption, efficiency, and congestion states tends to be impossible. Various publications in recent decades aimed to optimize individuals' routing decisions according to the 'path-finding problem'. Prior solutions replicated real-life behavior and aimed to reduce travel time (such as the Expected Value Model) \cite{Dantzig.1960} or, as in more recent publications, additionally increase travel time reliability \cite{U.S.FederalHighwayAdministration.2006}. Due to scale effects, the optimization of individual behavior does not necessarily result in an optimized overall state. Therefore, individual decisions need to be balanced over the collective to achieve an equilibrium state that admits a preferable routing solution for the individual while keeping the collectives system in an optimum at the same time. Traffic equilibrium models can be used and optimized for different purposes. Either they calculate the optimum routing solution for the target regions or they visualize the current real mobility behavior inside these regions. Individuals differ in their routing-decisions while traveling \cite{Zhou.2008} and so different equilibrium models exist according to the participants risk behavior: from risk-averse (TTB \cite{Uchida.}, PUE \cite{Lo.2006}), to risk-neutral (UE \cite{Wardrop.1952}, SUE\cite{CARLOSF.DAGANZO.1977}), and risk-prone (CVaR \cite{Rockafellar.2000}, RSUE \cite{Shao.2006}). 
The aforementioned models use resource consuming enumerating methods to estimate a global loss minimum. However Chen and Zhou \cite{Chen.2010} proposed a new lightweight method known as SOTA (Stochastic On-Time Arrival), which discretizes the problem into multiple independently solvable routing decisions. A further improvement to the SOTA model is the $\alpha$-reliable path model \cite{Chen.2005} that additionally characterizes routes according to their reliability with an $\alpha$-value. Commuters are then able to choose their shortest path based on their risk acceptance.
Finally, Zhou \cite{Zhou.2008} extends this model to create the so-called mean-excess traffic equilibrium (METE) model that allows to integrate real-time traffic information within the decision process and therefore also considers time-variant delays due to congestions' or other events. Considering the IOTB \cite{Nie.2006} algorithm the system's computational efficiency for routing decisions is drastically improved. As METE seems to improve commutation modeling, integrates stochastic measures of individual risk perception, is capable of incorporating real-time data, and enables short computation times, it is currently a favored candidate for the new mobility model described in the next chapter.

\section{Unified Mobility Estimation Model} \label{section:UMEM}

The main contribution of this paper is the presentation of a new mobility concept, which we call the 'Unified Mobility Estimation Model' (UMEM). This model presupposes assumptions of human behavior under the following conditions:
\begin{enumerate}
	\item[1.] Commuters
	\begin{enumerate}
	    \item[(a)] live in the focus areas and leave and return to their homes on average once a day.
	    \item[(b)] choose similar locations nearby for their daily routines if living in the same neighborhood.
	    \item[(c)] are familiar with their environment and have a set of daily and weekly routines.
	    \item[(d)] endeavor to optimize their travels regarding time and effort.
	\end{enumerate}
	\item[2.] Substantial daily routines, like going to work, going to school or shopping for food, clothes, etc. generate the majority of regional trips (excluding transit travel).
	\item[3.] Transit travel is not considered in the model but can be added subsequently.
	\item[4.] Inter-modal travel behavior is not considered in the model but can be added subsequently.
\end{enumerate}

Our model is based on the fundamental theory that mobility differences are mainly revealed by regional differences and rather less by social or ethical differences within these regions. In other words, the regional environment \& points of interests (POIs) adapt to a regional society's needs within a consolidated living area, and commutation is strongly impacted by it considering the aforementioned assumptions. This balancing phenomenon between supply and demand is well known in other domains like economics by the market equilibrium \cite{Applebaum.1961}. Consequently, to estimate the overall mobility in a region, it is sufficient to consider the regional infrastructure and distribution of residential homes considering the aforementioned mobility laws. Mobility metrics of the population are finally derived from our model without the need for simulating an individual’s mobility behavior.

\subsection{Step 1 - Target Area Delimitation \& POI Identification} \label{section:STP1_Target_Area}

In the first step of our model we identified living space clusters, as they offer a balanced geospatially delimited system with a working market equilibrium. Good sources for such an evaluation include building densities from Google Maps, satellite images, and data from the 'Global Human Settlement Layer' (GHSL) published by the European Union. In the next step we analyzed the density of important POI categories (supermarket, school, ...) inside the living space clusters $\vec{v_{LC}}$ that are limited by a boundary $\vec{v_{TA}}$ including an additional margin of the size $\vec{v_{PTE}}$ that reflects a 95\% probable chance that a traveler would move inside this boundary within one day according to the 'Physical Travel Energy (PTE) Law' from paragraph \ref{section:Physical_Travel_Energy}:

\begin{equation}
\begin{aligned}
\vec{v_{TA}}=\vec{v_{LC}}+\vec{\Delta v_{PTE}}=\left[\begin{matrix}x_{CR}\\y_{CR}\\\end{matrix}\right]+\left[\begin{matrix}\Delta x_{PCA}\\\Delta y_{PCA}\\\end{matrix}\right] \\
\forall\vec{\Delta v_{PTE}}:\int_{0}^{E\left(\left|\vec{\Delta v_{PTE}}\right|\right)}{P_{PTE}\left(2\cdot x_E\right)dx_E}<0.95
\end{aligned}
\end{equation}

where $P_{PCA}$ is the probability of expending $2\cdot x_E$ of energy over one day, $\vec{\Delta v_{PTE}}$ the vectorized additional distance from the boundaries of the living clusters and $E\left(\left |\vec {\Delta v_{PTE}}\right|\right)$ the expended commutation energy over this distance. The factor two of $x_E$ considers the doubled driving distance on returning back home. The boundary describes the most extreme case, in which a commuter travels to only one destination and back in one day.
Under the assumption that the quantity of POIs in the observed region correlates with individual consumer demands, by ranking the POIs according to their importance with respect to the demand $z_{POI}$ and taking into consideration the rank-size rule or Zipf’s law from paragraph \ref{section:Zipf_Law}, we derived the commutation probability $P_{POI}$ for different POI categories: 
\begin{equation} \label{eq:12}
P_{POI}\left(z_{POI} \right)=c_{POI}\cdot z_{POI}^{-\left(1+\frac{1}{\alpha_{POI}}\right)}
\end{equation}

where $c_{POI}$ and $\alpha_{POI}$ must be calibrated once by empirical data. We  assume that the ranks and calibration values in average do not vary much within a country, as the objects of demand, i.e. food, water, work, education and clothing are essentials of life and such demands can be described on a state or country level (for instance an European country has different demands to a country in Africa, but within each individual country, demands are likely to not differ as much as between countries). Due to the nature of the rank-size rule, the probability decreases considerably with each subsequent rank and therefore a few (approximately the first 10) important POIs are significant to the overall result.

\subsection{Step 2 - Weighted Potential Graph for Commutation} \label{section:STP2_Potential_Graph}

In the next step, the target areas need to be segmented into $z_{zo}$ quadratic zones with a side length of $l$. We slightly modified the 'Extended Radiation Model' mentioned in paragraph \ref{section:Radiation_Model} in line with the ranked POI distribution from Step 1, and generated a multidimensional geospatial weighted potential-graph for commutation (GWPC) based on equation \eqref{eq:6} by iterating through all zones and calculating the cumulative visitation probability $P_{ER}$ for the surrounding zones:

\begin{gather}
P_{ER}=\frac{P_>\left(n_{OZ}+s_{Track}\right)-P_>(n_{OZ}+n_{DZ}+s_{Track})}{P_>\left(n_{OZ}\right)}\notag\\
\langle P_>\left(x\right)\rangle=\frac{\frac{1}{1+x^\alpha}-\frac{1}{1+n_{Total}^\alpha}}{\frac{1}{1+n_{avg}^\alpha}-\frac{1}{1+n_{Total}^\alpha}},n_{Total}\geq x\geq n_{avg}\label{eq:13}\\
\alpha=\left(\frac{l}{36}\right)^{1.33},n_{avg}=\frac{n_{TA}}{z_{zo}}\notag\\
\forall\ n_{LZ},n_{DZ},n_{Total},s_{Track}: \sum_{Zones}\sum_{P_{POI}}{P_{POI}\cdot{n_{POI}}_{|Zone}}\notag
\end{gather}

where $n_{OZ}$ is the number of weighted POI opportunities (WPO) inside the origin zone, $n_{DZ}$ is the number of WPOs inside the destination zone, $s_{Track}$ is the weighted number of WPOs between the origin zone to the destination zone, and $n_{TA}$ is the number of weighted zones inside the target area. The WPOs depend on the daily importance of the POIs weighted by the $P_{POI}$ function from equation \eqref{eq:12}. 

\subsection{Step 3 - Trip Generation} \label{section:STP3_Trip_Generation}

As the GWPC only predicts the commutation probability of the different zones but does not describe travel behavior en route to these zones (e.g. if it is done in a single trip or in multiple trips throughout the day), we implemented further steps to describe mobility overall. In this step, we depicted commutation behavior by bringing the GWPC zones in relation with each other taking into consideration the Lévy flight step size probability and the 'Physical Travel Energy' distribution from paragraph \ref{section:Physical_Travel_Energy} and the 'Returners' \& 'Explorers' theory \& 'Human Mobility Motifs' theory from \ref{section:Returners_Explorers} \& \ref{section:Human_Mobility_Model}.
Starting from zones where commuters begin their daily travel (residential clusters), all zones are connected in steps to form a trip taking evolution strategies into consideration. The calculation process is discretized, and trips evolve at each step by adding a new random zone from the target area to the trip. At each evolution step, a trip-realisticity value $P_{Trip|Mod.}$ is calculated by considering all zones (here: $Zones$) the trip covers:

\begin{equation}
\begin{gathered}
P_{PTE,Mod.}=P_{PTE}\left(E\left(\sum_{Zones} d_{Zone_{i-1},Zone_i}\right)\right) \\
P_{motif}=P_{motif}(sk_{Zones})\\
P_{ER}=\prod_{i=1}^{Zones}P_{ER:Zone_{i-1},Zone_i} \\
P_{LF}=\prod_{i=1}^{Zones}{P_{LF}(d_{Zone_{i-1},Zone_i})} \\
P_{Trip|Mod.}=P_{PTE,Mod.}\cdot P_{Motif}\cdot P_{ER}\cdot P_{LF} \\
\end{gathered}
\end{equation}

where $d_{Zone_{i-1};Zone_i}$ is the Euclidean distance between two zones and $sk_{Zones}$ is the trip's exploration ratio according to \ref{section:Returners_Explorers}. In the current concept, the trip as a whole is unimodal and described for a specific modality mode (here: $Mod.$). Intermodality can be achieved by assigning modalities to different sections within a single trip. Unfinished trips share the same history with their preceding (evolution) steps and therefore probabilities that rate the commutation behavior, as defined by the 'Extended Radiation Model' $P_{ER}$ and by the 'Lévy Flight Model' $P_{LF}$, can be taken from the previous evolution step. Probabilities that rate the trip-realisticity defined by the 'Physical Travel Energy' constraint probability $P_{PTE,Mod.}$ and the 'Travel Motif' probability $P_{Motif}$ need to be recalculated at each evolution step. If an evolution step's trip-realisticity $P_{Trip|Mod.}$ falls below $P_{Trip,min}$, it is rejected, as it no longer appears sufficiently realistic. Considering the law of reoccurring returns from \ref{section:Social_Confined_Model} and the aforementioned constraints, commuters are forced to finish their trips at their homes daily.

\subsection{Step 4 - Travel Equilibrium \& Road Load Estimation} \label{section:STP4_Travel_Equilibrium}

\begin{figure}[b!]
\centerline{\includegraphics[width=9cm]{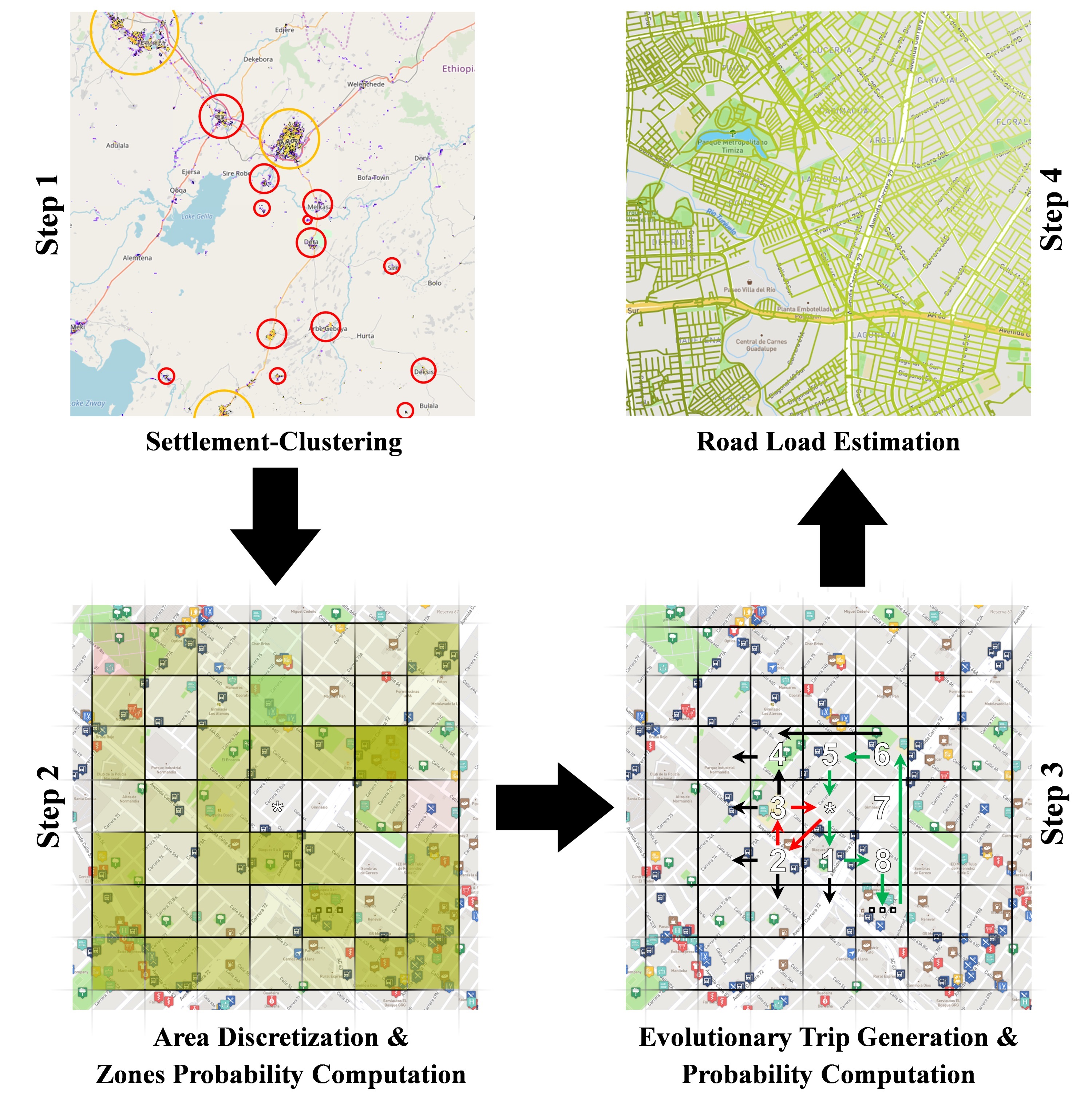}}
\caption{Basic steps of the Unified Mobility Estimation Model.}
\label{fig:4}
\end{figure}
\autoref{fig:4} illustrates the basic steps of the model. Using all valid trips and their corresponding probabilities $P_{Trip}$ from Step 3, we derived road loads constrained by a global traffic equilibrium. In its simplest representation a risk-neutral UE model (see Section \ref{section:Traffic_Equilibrium}) that only selects the shortest road connection between the origin zone and destination zones is chosen. More detailed models, such as METE can also be used with routing considerations. In all applied equilibrium models, the path flows $F_{Path}$ for each trip are calculated directly from the realisticity probability $P_{Trip|Mod.}$, which is normalized to the sum of realisticity probabilities $P_{Origin}$ of all trips that start from the origin zone. Combined with the proportion $\alpha_{Modality}$ of users using the selected modality and the population $n_{origin}$ in the origin zone, the path flow is calculated as follows:

\begin{equation}
\begin{aligned}
F_{Path}=\frac{P_{Trip|Mod.}}{P_{Origin}}\cdot\alpha_{Mod.}\cdot n_{origin} \\
P_{Origin}=\sum_{Trips_{Origin}}P_{Trip|Mod.}
\end{aligned}
\end{equation}

The final flow within a road segment $i$ of a specific road is then calculated by summing up all path flows covering the segment:

\begin{equation}
F_{Road,i}=\sum_{Trips_{Road,i}} F_{Path}
\end{equation}

The modality proportions $\alpha_{Modality}$ of the zones can initially underlie statistics of greater regions (e.g. country-level, state-level) and be gradually refined if more detailed data exists.

\section{Summary} \label{section:Summary}
In this paper, we proposed the Unified Mobility Estimation Model (UMEM), a novel model for simulating mobility. Our concept relies solely on spatial population data, points of interest, road networks, and the most essential daily needs that constitute the main reason for commuting within a region. The model combines several methods taken from recent mobility research, that share the aim of describing mobility on an abstract level rather than by studying the individuals' mobility behavior. One major advantage of the proposed model are the reduced data requirements, which might enable a worldwide prediction of mobility based on existing information. Furthermore, the model's parameters like the spatial resolution and the number of evolution steps are highly discretized, which allows adapting calculation expenses geospatially. In specific regions, the model can predict mobility with a high level of accuracy, while for other regions a lower degree of accuracy is chosen. This enables scalability, flexibility and performance. The UMEM model forms a theoretical foundation for future work. In subsequent studies, simulations, and real-world experiments, we will elaborate on how well adoptions and constraints adjust with most regions. We will demonstrate how to calibrate the model's parameters and analyze the required degree of detail for input data to obtain adequate results. Additionally, we will examine if other data sources with a higher resolution, such as GSM tracking data, will render the model's parameters even more robust and accurate for further predictions.

\section{Contributions} \label{section:Contributions}
David Ziegler initiated the idea of the paper, conceived the presented model, developed the theory \& theoretical formalism, and wrote this publication. Johannes Betz contributed to the research design, verified the methods and contributed to the final version of the publication. Markus Lienkamp made an essential contribution to the conception of the research project. He revised the paper critically for important intellectual content. Markus Lienkamp gave final approval of the version to be published and agrees to all aspects of the work. As a guarantor, he accepts responsibility for the overall integrity of the paper.

\printbibliography
\vspace{12pt}
\color{red}
\end{document}